# Comparative analysis of the lubrication performance of functionalized copolymers interacting with silicon, cobalt, and silver doped diamond-like carbon

Takeru Omiya[1, 2, 3*], Enrico Pedretti[4], Pooja Sharma[1], Albano Cavaleiro[1, 3], Arménio C. Serra[2], Jorge F. J. Coelho[2,3], Maria Clelia Righi[4*] and Fábio Ferreira[1, 3]

[1]*University of Coimbra, CEMMPRE, ARISE, Department of Mechanical Engineering, Rua Luís Reis Santos, 3030-788 Coimbra, Portugal;*

[2]*University of Coimbra, CEMMPRE, ARISE, Department of Chemical Engineering, Rua Silvio Lima, 3030-790 Coimbra, Portugal;*

[3]*Laboratory for Wear, Testing & Materials, Instituto Pedro Nunes, Rua Pedro Nunes, 3030-199 Coimbra, Portugal;*

[4]*Department of Physics and Astronomy, University of Bologna, Bologna, 40127 Italy.*

*Corresponding authors: takeru.omiya@student.dem.uc.pt, clelia.righi@unibo.it

## Abstract

This study examines the tribological behavior of diamond-like carbon (DLC) coatings doped with silicon (Si), cobalt (Co), or silver (Ag) in the presence of an amine-functionalized block copolymer lubricant. Under boundary lubrication, Si-doped DLC (Si-DLC) exhibited the lowest coefficient of friction (≈0.045) and nearly 45% lower wear than undoped DLC. Co-DLC showed moderate improvement, while Ag-DLC provided no significant benefit.

Cross-sectional FIB–TEM revealed thin tribofilms, 12–17 nm in thickness, on Si- and Co-doped surfaces. As reported for Si-DLC, these films incorporate copolymer-derived fragments, suggesting a similar composition for Co-DLC. These results indicate that dopant–polymer interactions are key to the development of self-organized boundary layers.

To gain atomic-level insight, first-principles calculations were carried out on the adsorption of the dimethylaminoethyl methacrylate (DMAEMA) unit, the copolymer's functional group. The calculated adsorption energies were –2.27 to –0.57 eV for Si-DLC, –1.73 to –1.49 eV for Co(0001), and –1.21 to –1.08 eV for Ag(111). The order of stability (Si > Co > Ag) was consistent with the experimental tribological ranking. Chemical bonding dominated for Si-DLC, while Ag showed mainly weak physisorption. Simulated pull-off forces further reflected this hierarchy, with N–Si bonds requiring about twice the force of N–Co and nearly five times that of N–Ag.

The correspondence between adsorption strength and tribological response highlights the decisive role of dopant species in tribofilm formation. These findings provide guidance for designing durable low-friction surfaces in applications such as electric drivetrains and precision mechanical systems.

## 1. Introduction

Polymer-based lubricants have long been utilized as viscosity index improvers and friction modifiers to enhance friction reduction and wear protection in various industrial applications, particularly in engine oils [1, 2]. These lubricants achieve surface protection and friction reduction through viscosity enhancement and the formation of adsorbed films on metal surfaces [3-6]. Among them, functionalized copolymers containing amine groups have demonstrated remarkable friction-reducing effects by forming lubricating films on steel surfaces [7-10]. For instance, Fan et al. reported that copolymers with amine groups form thick layers, effectively reducing friction on metal substrates [11].

Despite these advancements, most studies on functionalized copolymers have primarily focused on steel surfaces, leaving significant potential unexplored for their interaction with diamond-like carbon (DLC) coatings. DLC films are widely used in the automotive industry due to their high hardness, excellent wear resistance, and low friction properties [12-16]. However, the interactions between functionalized copolymers and DLC remain poorly understood, mainly because DLC's chemical inertness limits the adsorption of polymeric species onto its surface [14, 17]. To address this limitation, research on doped DLC has gained increasing attention, revealing that specific elements can modify surface reactivity and enhance interactions with lubricants [18-22]. Miyake et al. demonstrated that doping DLC with metals such as cobalt, cerium, magnesium, nickel, and titanium improves boundary lubrication properties under polyalphaolefin (PAO) lubricants [23]. Notably, Co-doped DLC exhibited the lowest friction coefficient under boundary lubrication using PAO with glycerol monooleate (GMO), highlighting its potential to enhance lubrication performance [23]. Furthermore, Khanmohammadi et al. investigated the tribological properties of tungsten and silver doped DLC coatings under water-glycol-based lubricants with ionic liquid additives, revealing that Ag-DLC exhibited superior mechanical properties and wear resistance [24].

While several studies have explored the combination of doped DLC with lubricants, investigations on their interaction with functionalized copolymers remain scarce. Recently, we demonstrated that silicon doping in DLC enhances tribofilm formation through N−Si bonding when combined with polymers containing dimethylaminoethyl methacrylate (DMAEMA) [25, 26]. However, no studies have examined the interaction between functionalized copolymers and metal-doped DLC coatings, leaving a significant gap in optimizing this novel lubrication system. Understanding how different dopants influence tribological performance and compatibility with functionalized copolymers is critical for further advancing this field.

Building on these findings, this study systematically investigates four types of DLC coatings, undoped DLC, Si-doped DLC (Si-DLC), Co-doped DLC (Co-DLC), and Ag-doped DLC (Ag-DLC), in combination with a functionalized polymer lubricant (PLMA-*b*-PDMAEMA). By comparing the friction and wear behavior of different dopants under identical experimental conditions, this study aims to elucidate how each dopant influences interactions with functionalized polymers. Moreover, a thorough computational investigation of how the adsorption mechanisms of the functional groups is affected by the different dopants is presented, highlighting the difference in bond strength, which determines the stability of the lubricant film in harsh tribological conditions. These findings not only expand the application potential of doped DLC in lubrication systems but also provide guidance on optimizing DLC surface chemistry to maximize its synergy with polymer-based additives.

## 2. Materials and methods

*2.1. DLC films*

In order to investigate the influence of dopants, four types of DLC films were prepared. These included an undoped DLC without any dopant and three DLCs doped with silicon, cobalt, and silver. All DLC films were deposited on AISI D2 steel substrates measuring 25 mm in diameter and 8 mm in thickness, with a surface roughness (Ra) of approximately 100 nm. The deposition was performed with sputtering methods, as outlined below.

Undoped DLC and Si-DLC films were produced using an unbalanced magnetron sputtering system (Teer Coatings Ltd) based on the procedure described in a previous study [25]. A chromium interlayer with a thickness of about 300 nm was first deposited to enhance the adhesion of the DLC film. For Si-DLC, the silicon content was controlled through different power applied on the graphite target and on the silicon target.

Co-DLC and Ag-DLC films were deposited using a high-power impulse magnetron sputtering (HiPIMS) Cyprium plasma generator (Zpulser Inc). Pure graphite targets with 99.95 percent purity and titanium targets with 99.99 percent purity, each measuring 150 mm by 150 mm by 10 mm, were prepared for the DLC films and titanium-based interlayers. Cobalt and silver pellets with a diameter of 10 mm and thickness of 2 mm were embedded into the graphite target to introduce cobalt and silver into the DLC films. The number of pellets was adjusted to control the dopant content. During all depositions, the steel substrates were placed at the center of the chamber and rotated at 23.5 revolutions per minute at a distance of 80 mm from each target. The carbon target was initially cleaned for 10 minutes at a pressure of 0.4 Pa using an average power of 600 W, in order to remove residues and oxides. The titanium target was similarly cleaned for 60 minutes at 0.35 Pa with a power of 250 W. A Ti interlayer was then deposited by direct current magnetron sputtering for 10 minutes at 0.3 Pa under a DC substrate bias of -60 V and a titanium target power of 1200 W in an argon plasma. This was followed by deposition of a TiN interlayer for 7 minutes at 0.3 Pa, with an argon gas flow ratio of 1 to 3. The Ti interlayer was approximately 300 nm thick and the TiN interlayer about 100 nm thick. Finally, DLC films were deposited by HiPIMS at a pressure of 0.4 Pa for 1 hour, with a micro-pulse duration of 6 microseconds, a micro-pulse period of 150 microseconds, and an overall pulse duration of 1800 microseconds. The substrate bias voltage was -80 V and the average power was 600 W.

All DLC films had a thickness of approximately 1 micrometer. The composition of the films was determined using scanning electron microscopy (SEM) equipped with energy-dispersive X-ray spectroscopy (EDS). Hardness and Young's modulus were measured using a NanoTest Vantage system (Micromaterials Ltd) with a Berkovich indenter. The nanoindentation tests were carried out with a maximum load of 3 mN, and the resulting indentation depth did not exceed 10% of the coating thickness (~1 μm). This ensured that the measured hardness and Young's modulus values reflect the intrinsic properties of the coating, minimizing the influence of the substrate. Measurements were performed at 16 locations to calculate the mean and standard deviation. In addition, Raman spectroscopy with a 532 nm laser was performed to determine the sp3 content of the films. Three spectra were recorded for each DLC film and averaged for statistical accuracy [27, 28]. **Table 1** summarizes the designations and the corresponding properties of these DLC films.

**Table 1. Properties of undoped DLC and doped DLCs.**

| Coating Name | Doping Element [at.%] | Roughness: $R_a$ [nm] | Hardness [GPa] | Young Modulus [GPa] | $sp^3$ content [%] |
|---|---|---|---|---|---|
| Undoped DLC | - | 4.9 ±0.1 | 15.6 ±0.6 | 167 ±3 | 26 |
| Si-DLC | Si: 8.3% | 6.4 ±0.1 | 12.4 ±0.5 | 165 ±3 | 42 |
| Co-DLC | Co: 10.7% | 7.8 ±0.4 | 18.4 ±1.7 | 191 ±3 | 13 |
| Ag-DLC | Ag: 7.6% | 7.0 ±0.9 | 14.0 ±0.5 | 142 ±2 | 24 |

*2.2. Lubricants*

The block copolymer of lauryl methacrylate (LMA) and 2-dimethylaminoethyl methacrylate (DMAEMA) was synthesized by Supplemental Activator and Reducing Agent Atom Transfer Radical Polymerization (SARA ATRP) (**Figure 1**) (See Reference [10] for detailed synthesis procedures). The amount of PDMAEMA (5 mol%) in the block copolymer was determined by $^1$H-NMR. The $M_n$ of the block copolymers was determined to be 38 kg/mol and Đ=1.08. Additionally, for tribological testing, 8 g of the copolymer was dissolved in 92 g of polyalphaolefin 4 (PAO4) to obtain a lubricant containing 8 wt.% PLMA-*b*-PDMAEMA.

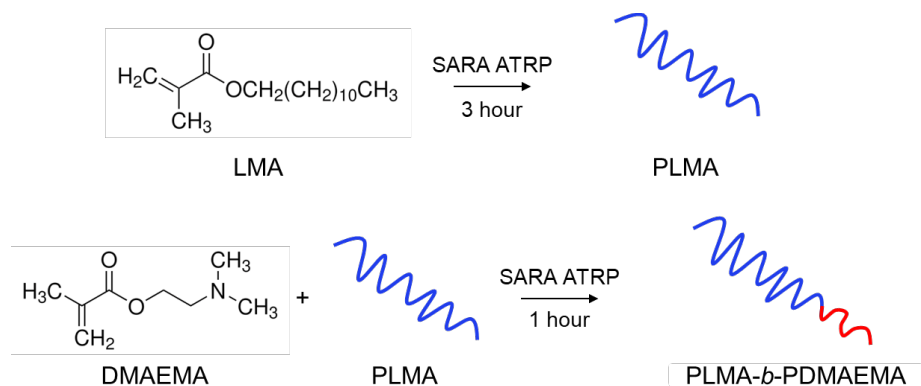

**Figure 1. Chemical structure of PLMA-*b*-PDMAEMA prepared by SARA ATRP.**

*2.3. Tribological tests*

Friction tests were conducted using a ball-on-disk tribometer (Rtec MFT-5000) in reciprocating motion to evaluate the friction performance of the DLC films. A new SiC ball, 3/8 inch in diameter with a surface roughness (Ra) of 25 nm, was used for each test. The tests were performed under a normal load of 3 N, corresponding to a maximum Hertzian contact pressure of 1.0 GPa. The friction tests were conducted at 80°C, as previous studies have demonstrated that this temperature is relevant for evaluating the friction reduction effect of functionalized polymers [5, 7]. Prior to each test, a running-in process was carried out for 5 minutes at a frequency of 3.3 Hz (corresponding to an average sliding velocity of approximately 0.1 m/s) to stabilize variations in the coefficient of friction (CoF). The friction test was applied with a frequency range from 5.0 to 0.2 Hz and a stroke length of 15 mm. The calculated λ values ranged between 0.3 and 0.03, indicating that the tests were conducted in the mixed to boundary lubrication regimes, as determined by the Hamrock-Dowson equation [29]. The steady-state CoF values were averaged for each sliding speed to assess the frictional behavior of the films.

Furthermore, wear tests were conducted using the same tribometer to evaluate the wear performance of the DLC films. The tests were performed at 80°C under a constant load of 100 N for a continuous sliding period of 20 minutes. A 3/8-inch diameter SiC ball was used as the counterpart, reciprocating over a stroke length of 1 mm at a frequency of 5 Hz.

After completing the tribological tests, the DLC samples were thoroughly cleaned with n-heptane to remove any residual debris. The wear profiles were then analyzed using white light interferometry (Rtec Instruments) to quantify the wear depth and volume. The wear rates were determined as the mean values obtained from three repeated tests, with standard deviations included to assess variability.

*2.4. Microscopic analysis and elemental composition*

Cross-sectional TEM lamellae were prepared using a Zeiss 550L cross-beam system. The wear track region of interest was first located using the in-situ SEM, and a protective platinum (Pt) layer was deposited on the surface by ion-beam induced deposition (IBID) to prevent damage from the $Ga^+$ ion beam during milling. Prior to FIB processing, a thin layer of gold (Au) had already been deposited on the sample surface as a conductive coating. Following Pt deposition, trenches were milled on both sides of the target region to define the lamella. Coarse milling was performed at high accelerating voltage (30 kV) and high beam currents, to obtain a 1μm thick lamella. The lamella was then lifted out using an in situ micromanipulator and attached to a Cu TEM grid using Pt deposition. Final thinning of the lamella to achieve electron transparency (~100 nm thickness) was also carried out at progressively lower energies (typically down to 5 kV or 2 kV) to reduce ion-beam damage, surface amorphization, and curtaining effect.

High-resolution transmission electron microscopy (HRTEM) and high-angle annular dark-field scanning transmission electron microscopy (HAADF-STEM) were performed using a Tecnai G2 F20 microscope operated at an accelerating voltage of 200 kV. This microscope was used to observe the cross-sectional structure of the wear track regions and to identify the layered morphology formed during tribological contact. To obtain higher spatial resolution and compositional information, complementary HAADF-STEM imaging, energy-dispersive X-ray spectroscopy (EDX) and electron energy loss spectroscopy (EELS) analyses were carried out using a Thermo Fisher Spectra 300 microscope operated at 300 kV, equipped with a SuperX detection system and a Gatan Continuum K3 EELS spectrometer with direct electron detection. Elemental maps for C and Si were collected to analyze the chemical distribution within the tribofilm and across the interface with the underlying Si-DLC layer.

*2.5. Computational method*

In order to study the interaction between the DMAEMA functional group of the copolymer and the dopants, we performed *ab initio* density functional theory (DFT) [30] calculations, which provide the required accuracy to describe chemical bonds between the molecule and the surfaces.

For Si-DLC, where Si atoms are carbide-forming and are present as substitutional in the DLC structure, we generated an amorphous surface model with 10 at. % Si concentration. In contrast, Co and Ag are known to form crystalline nanoparticles of a few nm size embedded in an amorphous DLC matrix. For Co-DLC, multiple studies report the detection of granular metallic particles (2.5 – 8 nm) in crystalline HCP phase dispersed in the disordered carbon matrix [31-33]. Similarly, silver in Ag-DLC does not form carbides, and is instead found in the form of crystalline nanoclusters with a size ranging 2 to 6 nm in the FCC phase [34-36]**.** Moreover, silver tends to undergo surface segregation over time through diffusion due to low miscibility and reduced surface energy [35], to form aggregates with sizes up to 20-100 nm at the DLC surface [36].

Since the size of these nanoparticles is much larger than the size of the DMAEMA monomers, the local interactions are more realistically modelled using periodic crystalline surfaces. The hcp(0001) surface was used to model Co nanoclusters, as it is the most stable at room temperature [37], while the fcc(111) surface was used for Ag.

First, we studied molecular adsorption of the DMAEMA monomers on Si-DLC, Co hcp(0001) and Ag fcc(111) surfaces to obtain the distribution of adsorption energies and observe the different bonding behaviors. Due to the large amount of local adsorption minima for molecules on surfaces, especially in the amorphous case, we performed a thorough sampling of the configuration space with the Xsorb program [38], which generates a large number of initial adsorption configurations and efficiently identifies the most stable ones by employing a two-step optimization process. First, many initial structures are generated by combining translations (adsorption sites) and rotations of the molecule. In the pre-optimization step, all these initial structures are optimized with large convergence thresholds on energy and forces to obtain a "rough" distribution of adsorption energies at a lowered computational cost, then a selected subset of configurations are fully optimized with more tight convergence thresholds ($10^{-3}$ Ry/bohr on forces and $10^{-4}$ Ry on energies).

On Co and Ag surfaces we considered the high-symmetry points (ontop, bridge and hollow/3-fold) as adsorption sites, and generated the initial adsorption structures with Xsorb by placing DMAEMA horizontally on the surface, with the N atom above each adsorption site. For the molecular rotations, we used the angles suggested by the symmetries of the crystalline surfaces (0°, 60° and 90°).

On Si-DLC, we instead considered the surface Si atoms as adsorption sites, positioning DMAEMA horizontally over the surface with the N atom of the amine group, and separately with the O atom of the carbonyl group, above each adsorption site.

Regarding molecular rotations, it was not possible to choose meaningful rotation angles from the symmetries of the surface due to its amorphous nature, and a reasonable tradeoff between computational cost and sampling coverage was obtained by on surface Si atoms (adsorption sites), including 4 horizontal rotations of multiples of 90° for each site.

In this way, by combining adsorption sites and molecular rotations, we generated 15 initial adsorption configurations for Co and Ag surfaces, all of which were fully optimized due to their low number, and 88 initial adsorption configurations on amorphous Si-DLC, of which 44 were fully optimized, choosing two configurations per site out of four for each of the two reference atoms in the molecule (N and O).

Using the most stable adsorption configuration for each substrate, we then tested the strength of the molecule-surface bonds by performing "pull-off" calculations, in which we progressively displaced the DMAEMA molecule by vertical steps of 0.25 Å. At each step, the z coordinate of one carbon atom of the molecule (on the opposite end to the amino group) was kept fixed, as well as the bottom layer of the slabs, while all other degrees of freedom were optimized. In this way it was possible to measure the pulling force at each displacement step as the z component of the residual force on that atom at the end of the optimization.

To gain more information on the nature and evolution of the bonds during the "pull off" process, we finally computed the charge density difference and projected density of states (PDOS) at a few selected steps. The charge density difference was calculated as the difference between the electronic charge density of the full structure (with the molecule adsorbed on the surface) and the charge densities of the molecule and slab alone (keeping the same atomic coordinates as in the full structure). The PDOS was obtained by projecting the electronic wavefunctions onto the atomic wavefunctions of the two atoms (N from the molecules and Si/Co/Ag from the surface) involved in the bond.

All *ab initio* calculations were performed with the plane waves DFT code from the Quantum Espresso package [39], using the Perdew-Burke-Ernzerhof (PBE) parametrization [40] of the generalized gradient approximation (GGA) for the exchange-correlation functional, and including spin-polarization. Ultrasoft pseudopotentials [41] were used to approximate the core electrons, with a kinetic energy cutoff of 40 Ry and charge density cutoff of 320 Ry. For the PDOS calculations, we employed the *projwfc* code of Quantum Espresso, projecting the wavefunctions onto orthogonalized atomic wavefunctions related to the atom pairs involved in the bonds, and calculating the *l-decomposed* density of states.

The Si-DLC slab was generated through melt-quench molecular dynamics simulations, following the same procedure from our previous work [26], obtaining a ~ 10 Å thick slab in a simulation cell of 18.8 Å × 15.0 Å x 24 Å (including the vacuum region above the surface to avoid interactions with periodic replicas). We used a 6-layers slab (~ 10 Å thick) to model the hcp-Co(0001) surface, in a simulation cell of 17.33 Å × 17.15 Å x 30 Å, and a 6-layers slab (~ 11 Å thick) for the fcc-Ag(111) model, in a simulation cell of 17.37 Å × 15.04 Å x 30 Å. The Brillouin zone was only sampled at the Γ-point due to the large size of the cells, after testing that higher k-points densities did not result in meaningful differences in the total energy and in the PDOS.

Since dispersion interactions are a fundamental part of molecular adsorption, we included dispersion corrections with the semi-empirical Grimme D3 scheme [42], which offers a good compromise between accuracy and computational cost. Since Grimme methods can sometimes result in over-binding with metallic surfaces, we tested it beforehand by calculating the surface energy of Co and Ag surfaces, which is a good indicator of surface reactivity. We obtaining 2.68 J/m$^2$ for Co(0001) and 1.26 J/m$^2$ for Ag(111), in good agreement with the experimental values of 2.51 J/m$^2$ (Co) and 1.24 J/m$^2$ (Ag) obtained by extrapolation from liquid surface tension [43], and computational values of 2.74 J/m$^2$ (Co) [44] and 1.16 J/m$^2$ (Ag) [45] obtained with more accurate methods.

Table 2. Surface energy (J/m$^2$) for Co, Ag and diamond surfaces calculated with Grimme DFT-D3 and DFT-D2, with reference experimental values and computational results calculated with higher levels of theory.

|  | DFT-D3 | DFT-D2 | Exp. Ref. [43] | Comp. Ref. |
| --- | --- | --- | --- | --- |
| Co(0001) | 2.68 | 3.04 | 2.51 | 2.74 [44] |
| Ag(111) | 1.26 | 1.39 | 1.24 | 1.16 [45] |
| C(001) | 4.98 | - | - | 4.83 [46] |

For comparison, the Grimme D2 [47] scheme resulted in an overestimation of surface energy, reaching 3.04 J/m$^2$ for Co and 1.39 J/m$^2$ for Ag. Finally, we calculated also the surface energy of the crystalline 2x1 dimer-reconstructed C(001) surface of diamond to assess the accuracy in describing diamond-like materials, obtaining a value of 4.98 J/m$^2$, in good agreement with the reference computational result of 4.83 J/m$^2$ that employed hybrid functionals [46]. From these tests we concluded that the Grimme-D3 scheme is suitable for describing with reasonable accuracy all our systems.

## 3. Results and Discussion

### 3.1. Tribological test

The influence of different dopants on the frictional behavior was evaluated, and the coefficients of friction (CoF) at various sliding speeds are presented in **Figure 2**. Undoped DLC and Ag-DLC exhibited the highest CoF values, with Ag-DLC showing a slightly higher CoF than undoped DLC in the low-speed region. In contrast, Si-DLC consistently demonstrated the lowest CoF across the entire speed range. Co-DLC showed a comparable CoF to that of Si-DLC at higher speeds; however, its CoF gradually increased as the sliding speed decreased. To further examine the frictional characteristics under low-speed conditions, where the differences among samples were more evident, additional tests were performed at a sliding speed of 0.2 Hz (0.006 m/s) for a duration of 40 minutes. All other testing parameters, including load and temperature, were kept the same as in the previous measurements. The results of these additional tests are presented in **Figure S1** of the Supplementary Information, and the observed trend in CoF values was consistent with that shown in **Figure 2**.

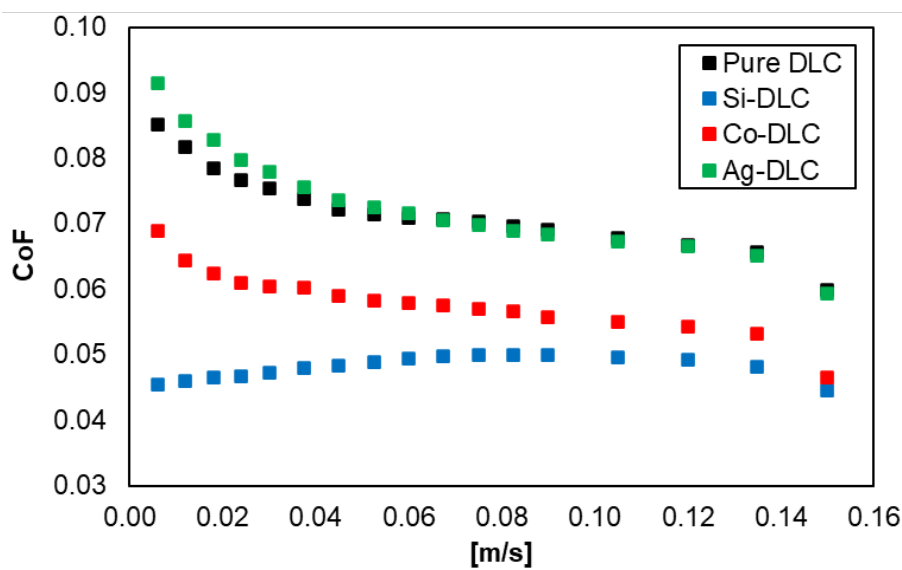

**Figure 2. Comparison of friction performance between undoped DLC and various doped DLC in the presence of PLMA-*b*-PDMAEMA.**

Next, a wear test was conducted to compare the wear resistance of each DLC film. **Figure 3(a)** shows the average wear rate and standard deviation for each sample based on three tests. The results indicate that undoped DLC exhibited the highest wear rate, revealing that wear was suppressed by the dopants. However, no significant differences among the dopants were observed in the wear test. Additionally, the evolution of CoF during the wear test is shown in **Figure 3(b)**. The dark lines represent the average CoF, and the light-colored areas represent the standard deviation at each time point. Unlike the wear rate results, differences among the dopants were observed in the CoF comparison. Similar to the results in **Figure 2**, Si-DLC exhibited the lowest CoF. Similarly, the results of wear tests using PLMA alone, which lacks the DMAEMA functional group of the functionalized copolymer, are provided in **Figure S2** of the Supplementary Information. While the addition of DMAEMA had no noticeable effect on undoped DLC, it clearly influenced both friction and wear behavior in all three types of doped DLC films, indicating that the presence of DMAEMA enhances tribological performance when combined with dopants.

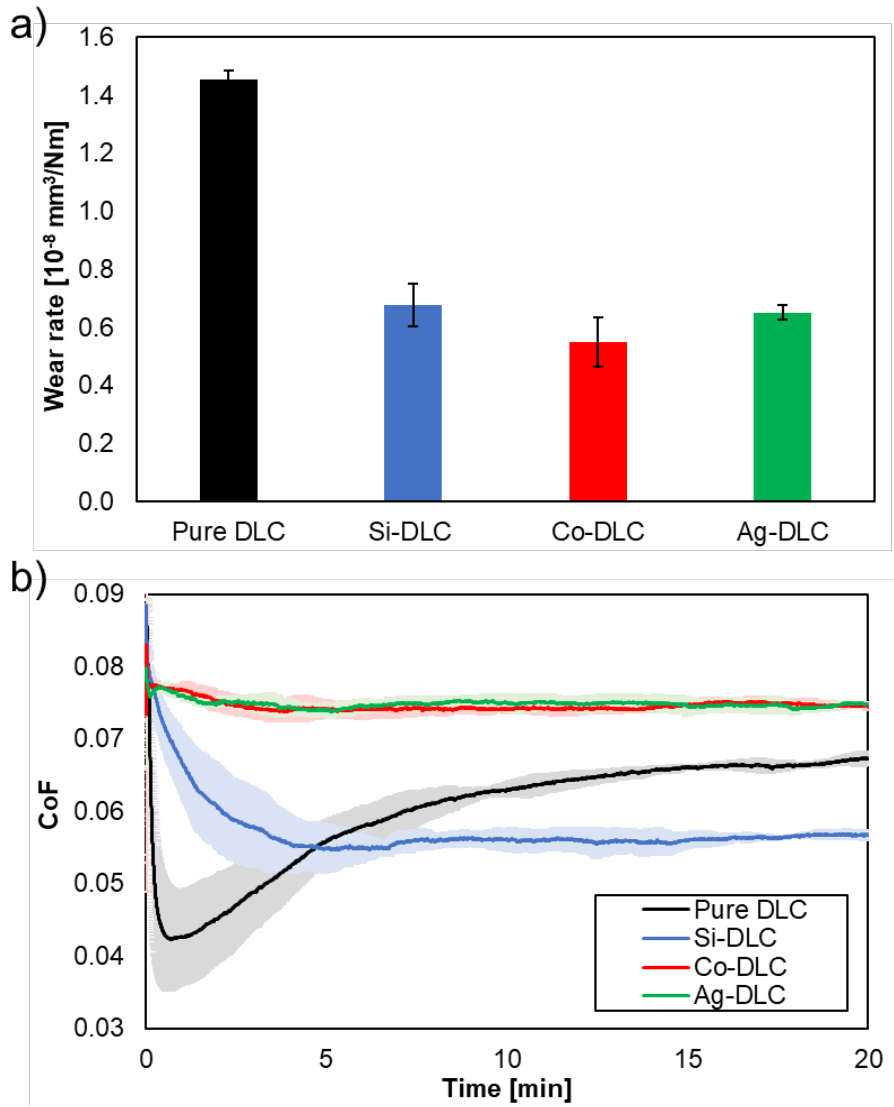

**Figure 3. Comparison of wear resistance performance between undoped DLC and various doped DLC in the presence of PLMA-b-PDMAEMA (a) Wear rate (b) CoF behavior during wear test.**

## 3.2. Characterization of tribofilm morphology

To investigate the structure of the tribofilm, FIB-TEM analysis was conducted using the Si-DLC and Co-DLC coatings, which showed the most significant effects in the previous friction tests. Focusing on the low-speed region where the most pronounced effects were observed, additional friction tests were performed for 40 minutes at a speed of 0.2 Hz (0.006 m/s). All other parameters, including load and temperature, were kept the same as in the previous friction tests. The corresponding friction test results are shown in **Figure S3** of the Supplementary Information.

To elucidate the morphology and thickness of the tribofilm formed in the presence of PLMA-*b*-PDMAEMA, cross-sectional FIB-TEM analysis were conducted on both Si-DLC and Co-DLC coatings. **Figure 4(a)** presents a representative HAADF image of the Si-DLC sample, while **Figure 4(b)** shows that of the Co-DLC sample. In both images, a protective Pt layer was deposited on top of a thin Au film to prevent damage to the tribofilm during subsequent Pt deposition. The Au layer was applied in advance to further protect the tribofilm structure. Beneath these protective layers, a clearly defined tribofilm was observed on both the Si-DLC and Co-DLC coatings, with thicknesses of approximately a dozen nanometers.

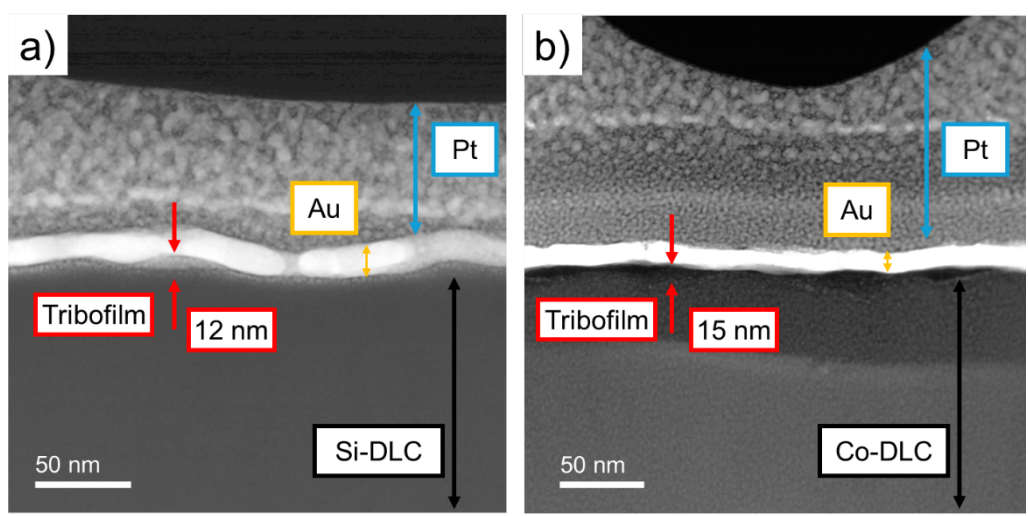

**Figure 4. Cross-sectional HAADF images of the tribofilms formed on (a) Si-DLC and (b) Co-DLC after sliding in the presence of PLMA-*b*-PDMAEMA. A thin Au layer and a protective Pt layer were deposited to prevent damage during sample preparation.**

**Figure 5** shows HAADF images and elemental maps for the Si-DLC and Co-

DLC samples. **Figure 5(a)** displays the HAADF image of the Si-DLC sample. **Figure 5(b)** presents an overlay of carbon (red) and silicon (blue) maps, revealing the formation of a carbon-rich tribofilm atop the silicon-containing DLC layer. Similarly, **Figure 5(c)** shows the HAADF image of the Co-DLC sample, and **Figure 5(d)** overlays carbon (red) and cobalt (blue) maps, indicating the presence of a carbon-rich tribofilm above the cobalt-containing DLC layer. These observations confirm that tribofilms derived from PLMA-*b*-PDMAEMA form on both Si-DLC and Co-DLC surfaces. In a previous study, the formation of tribofilms on Si-DLC in combination with functionalized copolymers was directly confirmed using X-ray Photoelectron Spectroscopy (XPS) and Time-of-Flight Secondary Ion Mass Spectrometry (ToF-SIMS) analyses [26, 48]. Based on the similar tribofilm morphology observed here for Co-DLC, it is reasonable to infer that a comparable tribofilm formation mechanism is also active in this case. The results support the interpretation that the amine-containing segments of PLMA-*b*-PDMAEMA anchor to the doped DLC surfaces, while the alkyl segments contribute to the formation of a low-friction surface layer.

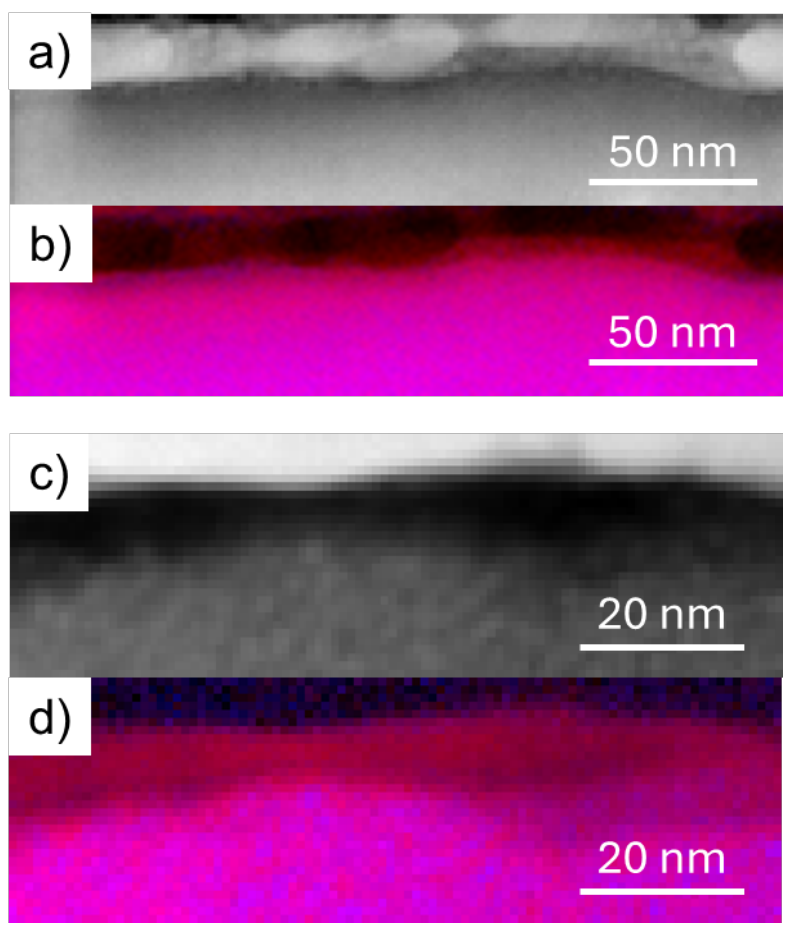

**Figure 5. HAADF images and EDX elemental maps of the tribofilms formed on Si-DLC and Co-DLC. (a) HAADF image of the Si-DLC sample; (b) overlay of carbon (red) and silicon (blue) maps showing a carbon-rich tribofilm above the silicon-containing DLC layer; (c) HAADF image of the Co-DLC sample; (d) overlay of carbon (red) and cobalt (blue) maps showing a carbon-rich tribofilm above the cobalt-containing DLC layer.**

### 3.3. First principles calculations

The main pre-requisite for a lubrication system that performs well in the severe conditions of boundary lubrication is good adhesion between the additives and the tribological surface, which enables the formation of a stable adsorbed film capable of withstanding harsh mechanical stresses. In the context of exploiting the synergistic effects of functionalized copolymers and doped DLC coating, the effectiveness of the combination is dependent on the strength of the interactions between the functional groups of the copolymers and the dopant atoms. Therefore, we carried out a comparative computational study to investigate molecular adsorption of the DMAEMA functional group of PLMA-b-PDMAEMA on Si-, Co- and Ag-doped DLC.

For the Si-DLC model we employed an amorphous structure with substitutional Si, while for Co-DLC and Ag-DLC we considered the Co(0001) and Ag(111) crystalline surfaces. This choice is justified by the fact that while carbide-forming Si atoms are randomly positioned within the amorphous carbon structure, Co and Ag aggregate to form crystalline nanoparticles of a few nm size (a more detailed discussion can be found in the Methods section). Since these nanoparticles are much larger than the DMAEMA monomers, most of the local molecule-dopant interactions occur on crystalline surface regions.

We carried out the molecular adsorption study with the program Xsorb, automatically generating a large number of initial adsorption configurations, which were then partially optimized with large convergence thresholds, and only a subset of the configurations with lower energy were fully optimized, to reduce the computational cost of this screening procedure (more information in the Methods).

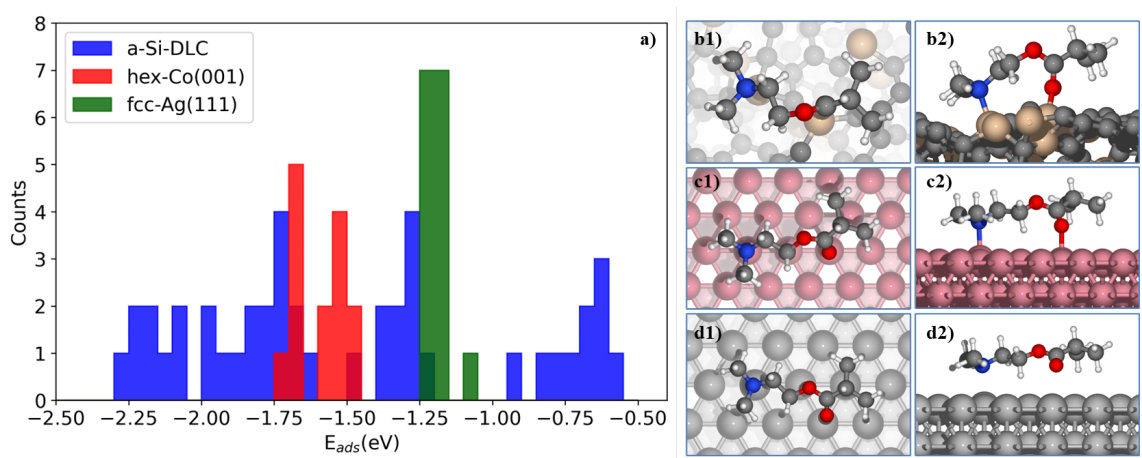

**Figure 6. Adsorption of DMAEMA on Si-DLC, Co(0001) and Ag(111). a) histogram of the**

**adsorption energy distribution on the surfaces for the three dopants; b-d): top(1) and lateral(2) view of a representative adsorption configuration on Si-DLC (b), Co (c) and Ag (d) surfaces. Si-DLC exhibits the strongest adsorption, with a notable number of configurations below -2 eV, followed by Co, and finally Ag, for which adsorption is the weakest. In all cases the molecule-surface interactions involve the N atom in the amine group, and often the O atom in the carbonyl group. C atoms are colored in grey, O in red, N in blue, H in white, Si in yellow, Co in pink, and Ag in silver.**

The adsorption energies for DMAEMA on Si-DLC, Co(0001) and Ag(111), defined as the difference between the energy of the adsorbed configuration and those of the isolated slab and molecule, i.e. $E_{ads} = E_{slab+mol} - (E_{slab} + E_{mol})$, are reported in **Figure 6**. As visible in the adsorption energy distribution histogram in **Figure 6(a)**, the ranges for adsorption energies are [-2.27, -0.57] eV for Si-DLC, [-1.73, -1.49] eV for Co(0001), and [-1.21, -1.08] eV for Ag(111). Expectedly, the crystalline surfaces exhibit much narrower energy distributions, while Si-DLC presents a significantly broader range due to the larger number of local environments that result from its amorphous structure, including a few physisorbed configurations. Despite this fact, if we look at the lowest-energy configurations for each surface, Si-DLC provides the strongest adsorption, followed by Co, and finally Ag. Qualitative information on the nature of the interactions (chemical bonding or dispersion interactions) can be obtained by extracting the dispersion contribution to adsorption energy, which is in the range of [-0.94, -0.57] eV for Si, [-1.41, -1.32] eV for Co and [-1.13, -1.06] eV for Ag. This suggests that dispersion interactions play a minor role in adsorption on Si-DLC, which involve mostly chemical bonding, while they play a more important role for Co and are almost the only contribution for Ag, which is only weakly physisorbed.

This trend in adsorption energies and the dispersion contribution considerations correlate well with the experimentally measured CoF values in **Figure 2**, with Si-DLC showing a significant friction reduction, larger than Co-DLC, and Ag-DLC having no improvement compared to undoped DLC.

To better understand the strength of the bonds between the nitrogen atom in DMAEMA and the dopants, which is relevant for the ability of the lubricant film to withstand the action of mechanical forces during sliding, we measured the resistive force to the progressive vertical displacement (pulling) of the molecule. The peaks in resistive force can be interpreted as the minimum instantaneous mechanical force that is required to detach the molecule from the surface and is a key aspect for the stability of the tribofilm in boundary lubrication conditions.

The resistive force trends are reported in **Figure 7**, along with energy variation with respect to the initial adsorption minimum, for the most stable adsorption configuration for each substrate. By comparing resistive forces just prior to the bond breaking between the nitrogen atom of the molecule and the surface, a clear difference can be seen between Si and metals, with a much sharper peak of ~1.25 eV/Å for Si, double than Co (~0.75 eV/Å) and five times larger than Ag (~0.25 eV/Å). Notably, Si-DLC also presents a high force peak related to O-Si bond breaking. Ag, on the other hand, show no clearly distinguishable force peaks, corroborating the hypothesis that no chemical bonding (to be broken) are formed, and the small resistive force is mainly from dispersion interactions.

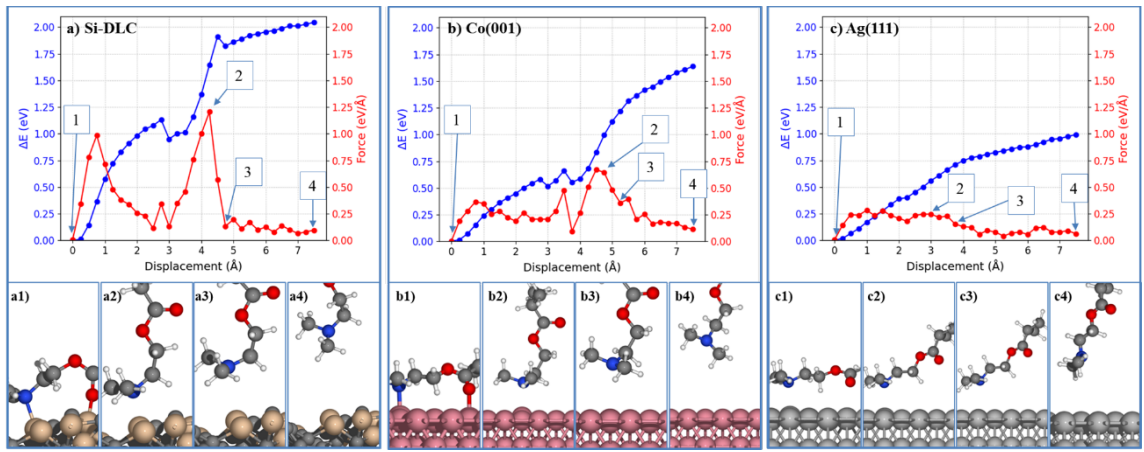

**Figure 7.** Resistive force to vertical displacement for DMAEMA adsorbed on Si-DLC (a), Co(0001) (b) and Ag(111) (c) surfaces, along with the energy variation with respect to the adsorption minimum. Snapshots during the pulling of the molecule are presented in (a1-a4), (b1-b4), (c1-c4) for Si, Co and Ag respectively. The numbers in the atomistic images correspond to the numbers in the graphs, associated to different moments: 1) initial adsorption minimum, 2) force peak just before bond breaking, 3) force drop after bond breaking, 4) full separation. C atoms are colored in grey, O in red, N in blue, H in white, Si in yellow, Co in pink, and Ag in silver. The moment just before N-Si bond breaking is labelled by (2), and corresponds to a high resistive force for Si, lower (half) for Co, and almost negligible for Ag, which does not exhibit clear peaks and the resistive force is almost uniform during the pulling process, indicating that no strong chemical bonds to break were present. Si-DLC also present a high force peak related to O-Si bond breaking, which is found (significantly smaller) also in Co.

The fact that for Si-DLC the difference in the instantaneous force required to break bonds is larger than the difference in adsorption energies with Co and Ag could partially explain the experimental CoF behavior of the wear test in **Figure 3(b)**, which is carried out at significantly higher load than the previous friction test, as well as the different trend of the friction tests itself at low speeds in **Figure 2**, where Si-DLC displays a consistently low CoF, while Co (and Ag) undergo a friction increase when transitioning towards the boundary lubrication regime. All these aspects considered together seem to suggest that the resistive forces to bond breaking in Si-DLC are large enough to form a tribofilm stably anchored to DLC surface, while for metals this is not the case, with Co-DLC forming a tribofilm, but weakly anchored, and no tribofilm formation at all for Ag-DLC.

To elucidate better the nature of the bonds and the reason behind the differences in adsorption energies and resistive forces for the three dopants, we calculated the charge density difference during the pulling process, which indicates the amount of electronic charge that is displaced between the molecule and surface when they are in contact. The charge density difference isosurface during the pulling process of DMAEMA on the three coatings are presented in **Figure 8**, where it's evident that for Si-DLC and Co a significant charge transfer occurs during adsorption, indicating the formation of chemical bonds, which are then broken as the distance increases, with the concurrent disappearing of the associated charge transfer. Notably, the charge transfer for Si-DLC is larger than for Co, with additional transfer on the surface, confirming what we observed in our previous work on the Si-doped crystalline diamond C(001) surface [25]. For Ag, instead, no relevant charge transfer is observed, which confirms the mostly dispersive nature of adsorption which was inferred from the analysis of the van der Waals contribution to adsorption energy, and explaining why no resistive force peaks are measured in the pulling process, as there are no "discrete" chemical bonds to be broken.

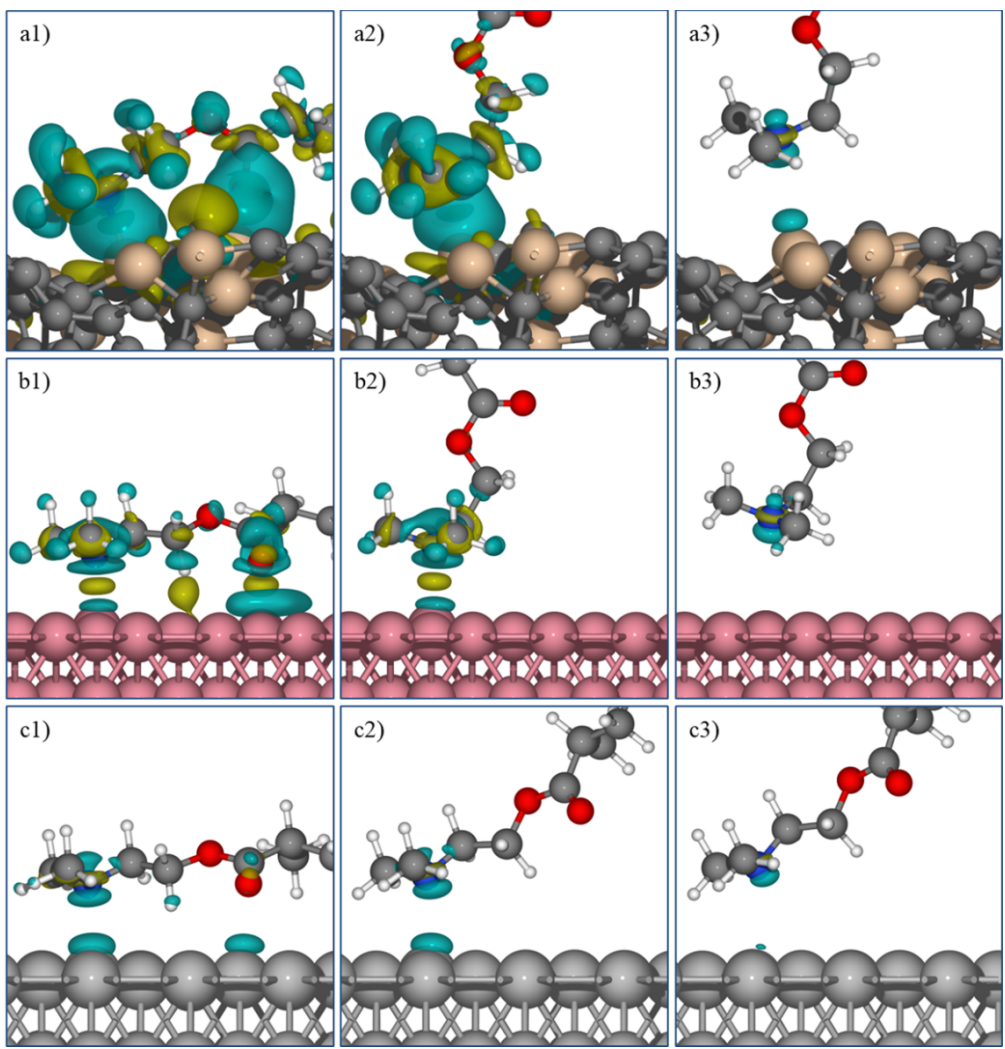

**Figure 8.** Charge density difference isosurfaces (0.002 e/Bohr$^3$) during the vertical pulling process for DMAEMA on Si-DLC (a1-a3), Co(0001) (b1-b3) and Ag(111) (c1-c3). The snapshots correspond to (1) initial adsorption minimum, (2) force peak just before bond breaking, (3) force drop after bond breaking, as indicated by the label in Figure 7. Charge accumulation is colored in yellow, depletion in turquoise. C atoms are colored in grey, O in red, N in blue, H in white, Si in yellow, Co in pink, and Ag in silver. Both Si-DLC and Co show a significant charge redistribution, with accumulation in the midpoint of N-Si/Co and O-Si/Co bonds, which suddenly disappears in 3), indicating the presence of chemical bonds that are broken by increasing the distance. In particular, Si-DLC shows the largest charge redistribution, with additional charge transfer on the surface. For Ag, instead, the charge displacement is almost negligible, suggesting a mostly non-chemical nature of the adsorption.

Finally, we computed the projected electronic density of states (PDOS) on the N(2p) orbitals of nitrogen in DMAEMA and Si(2p), Co(3d) and Ag(3d) of the surface atoms involved in the bond. The variation of the PDOS during the bond breaking is reported in **Figure 9**.

For reference, in all cases at large distances (a4, b4, c4 in the figure) the N(2p) state is present just below the Fermi level, corresponding to the non-bonded nitrogen lone pair. For Si-DLC even a Si dangling bond can be observed as a state just below Fermi level.

At the initial adsorption minimum (a1, b1, c1 in the figure) a major difference can be seen across the three dopant species: for Si-DLC, a notable change can be observed in the PDOS compared to large distance, with the N(2p) lone pair state and Si(2p) dangling bond that disappear, moving towards lower energies, thus indicating the formation of a chemical bond. The same also happens for Co, albeit on a smaller scale, involving mostly N(2p) but only minimally Co(3d), while for Ag the N(2p) state undergoes a minor variation, with a small shift to lower energy, confirming that adsorption is dominated by dispersion interactions and not covalent charge transfer. Moreover, for Si and Co the before (1-2) and after (3-4) bond breaking features can be clearly distinguished, with a sharp transition, while for Ag the variation consists in a slow migration of the N(2p) state towards the Fermi level as the molecule is vertically displaced, compatible with the picture that electrons are not significantly involved in the adsorption/desorption process.

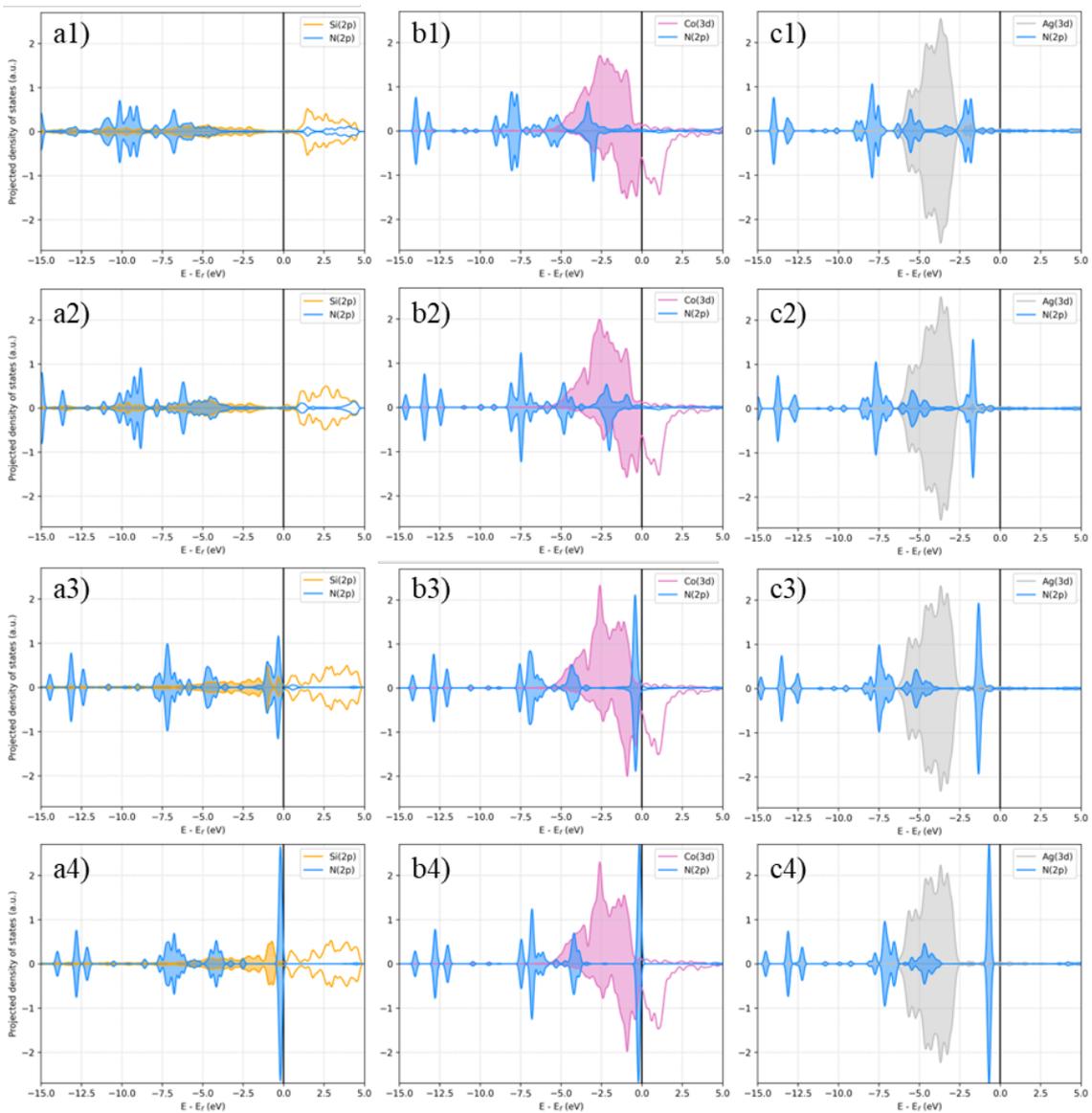

**Figure 9.** Projected electronic density of states (PDOS) during the vertical displacement for DMAEMA on Si-DLC (a), Co(0001) (b) and Ag(111) (c). The labels (1-4) correspond to the snapshots from Figure 7. The zero of the energy is set to the Fermi level. The graphs show the PDOS on the N(2p) of DMAEMA and the Si(2p), Co(3d) and Ag(3d) of the surface atom involved in the bonds. At large distance (a4-c4) the N lone pair is visible as a sharp peak just below the Fermi level. For Si (and in smaller measure Co), the lone pair states disappear when the molecule is adsorbed, with the N(2p) states moving towards low energy, indicating the formation of a chemical bond, while for Ag only a small shift to lower energy is observed, confirming the mostly dispersive nature of the molecule-surface interaction.

This computational analysis of adsorption of the functional group on the three differently doped DLCs indicates a clear trend in the formation and stability of bonds, which correlates well with the experimentally measured tribological performance. Other mechanisms may play a role in determining tribological performance and are the objective of further investigation, in particular the surface distribution of dopant atoms, which provide a uniform availability of adsorption sites for Si-DLC, while for Co-DLC and Ag-DLC they are concentrated in local regions corresponding to the crystalline nanoparticles embedded in the DLC matrix. However, the strength of the interaction between the functional group and the doped surface is always a necessary feature to obtain a stable tribofilm that performs well even in boundary lubrication regime, and this analysis of the chemical bonding aims to indicate a direction for the research of new lubrication systems that exploit the synergistic effects of additive functionalization and DLC doping.

## 4. Conclusion

In this study, DLC coatings doped with silicon, cobalt, or silver were evaluated in combination with an amine-functionalized block copolymer (PLMA-*b*-PDMAEMA) to elucidate the role of dopants in tribofilm formation and friction reduction. Tribological testing revealed that Si-doped DLC exhibited the lowest friction in boundary lubrication regimes, followed by Co-doped DLC, while Ag-doped DLC showed no improvement compared to the undoped counterpart.

Cross-sectional FIB–TEM and EDX analyses confirmed the presence of uniform carbon-rich tribofilms on Si- and Co-doped surfaces, suggesting that anchoring occurs via functional groups, with hydrocarbon segments contributing to low shear and load-bearing properties. First-principles calculations supported these findings, showing a clear trend in adsorption strength (Si > Co > Ag), consistent with the tribological results. Pull-off force simulations indicated that N–Si bonds require approximately twice the force to break compared to N–Co, and five times more than N–Ag. Charge density and projected density of states (PDOS) analyses revealed covalent-like bonding for Si and Co, in contrast to weak physisorption for Ag.

These results demonstrate that the formation of robust, low-friction tribofilms depends strongly on the chemical affinity between dopants and lubricant functional groups. This insight offers a practical guideline for developing advanced lubrication systems: selecting dopant elements, such as silicon, that promote chemical anchoring of functionalized additives to form durable boundary films.

**ACKNOWLEDGMENTS**
This research is sponsored by national funds through FCT – Fundação para a Ciência e a Tecnologia, under the projects UID/00285 - Centre for Mechanical Engineering, Materials and Processes, LA/P/0112/2020, 2023.08138.CEECIND/CP2832/CT0005, SmartHyLub (2022.05603.PTDC), iLub (2022.15609.UTA), UniLub (2023.17357.ICDT) and by the Taiho Kogyo Tribology Research Foundation (Grant No. 22A25). This project has received funding from the European Union's Horizon 2020 research and innovation programme under grant agreement No 101007417, having benefited from the access provided by ALBA Synchrotron, Barcelona; Karlsruhe Institute of Technology, Karlsruhe; Consejo Superior de Investigaciones Científicas - Centro Nacional de Microelectrónica, Barcelona; Fundacio Institut Català de Nanociència i Nanotecnologia, Barcelona within the framework of the NFFA-Europe Pilot Transnational Access Activity, proposal ID580.

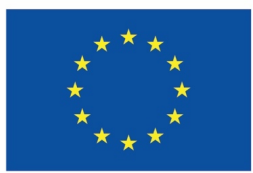

These results are part of the "Advancing Solid Interface and Lubricants by First Principles Material Design (SLIDE)" project that has received funding from the European Research Council (ERC) under the European Union's Horizon 2020 research and innovation program (Grant agreement No. 865633).

**Declarations**

**Conflict of interest:** The authors have no competing interests.